\documentclass[aps,prl,twocolumn,epsfig,floatfix]{revtex4}

\usepackage{graphicx}
\usepackage{amssymb,amsmath,amsthm,amsfonts}

\newcommand{\beq}{\begin{equation}}
\newcommand{\eeq}{\end{equation}}
\newcommand{\bdis}{\begin{displaymath}}
\newcommand{\edis}{\end{displaymath}}
\newcommand{\bea}{\begin{eqnarray}}
\newcommand{\eea}{\end{eqnarray}}
\newcommand{\barr}{\begin{array}}
\newcommand{\earr}{\end{array}}
\newcommand{\bfig}{\begin{figure}[!]}
\newcommand{\efig}{\end{figure}}

\newcommand {\gM}{g_{_\mathit{\! M}}}
\newcommand {\gtilde}{\tilde{g}}
\newcommand {\gtildeM}{\tilde{g}_{_\mathit{\! M}}}

\newcommand {\hvecm}{\mathbf{h}_{_\mathit{\! M}}}

\newcommand {\Msat}{M_s}

\newcommand {\mvec}{\mathbf{m}}

\newcommand {\evecx}{\mathbf{e}_{x}}
\newcommand {\evecy}{\mathbf{e}_{y}}
\newcommand {\evecz}{\mathbf{e}_{z}}

\begin{document}

\title{Conservative effects in spin-transfer-driven magnetization dynamics}

\author{G. Bertotti,$^{1}$ C. Serpico,$^{2}$ I.D. Mayergoyz,$^{3}$}
\affiliation{
$^{1}$INRIM, Istituto Nazionale di Ricerca Metrologica, Strada delle Cacce 91, 10135 Torino, Italy \\
$^{2}$Department of Electrical Engineering, University of Naples "Federico II", via Claudio 21, 80125 Napoli, Italy \\
$^{3}$Department of Electrical and Computer Engineering, University of Maryland, College Park MD 20742}

\received{\today}

\begin{abstract}
It is shown that under appropriate conditions spin-transfer-driven magnetization dynamics in a single-domain nanomagnet is conservative in nature and admits a specific integral of motion, which is reduced to the usual magnetic energy when the spin current goes to zero. The existence of this conservation law is connected to the symmetry properties of the dynamics under simultaneous inversion of magnetisation and time. When one applies an external magnetic field parallel to the spin polarization, the dynamics is transformed from conservative into dissipative. More precisely, it is demonstrated that there exists a state function such that the field induces a monotone relaxation of this function toward its minima or maxima, depending on the field orientation. These results hold in the absence of intrinsic damping effects. When intrinsic damping is included in the description, a competition arises between field-induced and damping-induced relaxations, which leads to the appearance of limit cycles, that is, of magnetization self-oscillations.
\end{abstract}


\maketitle

The spin-transfer phenomenon and related spintronic applications have been the focus of considerable research in the past two decades \cite{Slonczewski1996,Berger1996,Tserkovnyak2005,Haney2008,Brataas2012}. This research has been dominated by experimental and theoretical studies of spin-transfer-induced magnetization switching \cite{Katine2000,Bazaliy2004,Kubota2008,Liu2012}, as well as spin-transfer-driven magnetization self-oscillations \cite{Kiselev2003,Rippard2004,Krivorotov2005,Bertotti2005,Bertotti2009}. These studies have all been based on the seed idea that spin transfer manifests itself as a non-conservative torque that competes with intrinsic (thermal) damping. In particular, it has been realized that the mutual compensation of non-conservative effects caused by spin transfer and thermal damping is the physical mechanism for the formation of magnetization self-oscillations \cite{Slonczewski1996,Bertotti2005,Li2005}.

It is demonstrated in this Letter that in single-domain nanomagnets spin transfer may act as a purely conservative torque when electron spin polarization is directed along the intermediate (i. e., hard in-plane) anisotropy axis. Under these conditions, the following new physical features emerge: the appearance of purely conservative magnetization dynamics with closed precession-type trajectories; the existence of a special integral of motion for this conservative dynamics, which is reduced to the conventional
magnetic energy at zero spin current; a very unique global bifurcation in magnetization dynamics occurring at a specific critical value of the injected spin-polarized current; the conversion of the conservative dynamics into monotone relaxation when an in-plane dc magnetic field is applied along the intermediate anisotropy axis; the existence of a Lyapunov function governing these field-induced relaxations as well as the appearance of field-induced interlacing of the basins of attractions of the critical points of the dynamics. The origin of all these new physical features can be traced back to the special symmetry of magnetization dynamics appearing in the case
when both electron spin polarization and applied dc magnetic field are directed along the intermediate axis of magnetic anisotropy.

The described new physical features of magnetization dynamics appear when intrinsic damping effects are neglected. When these damping effects are accounted for, the mutual compensation of the non-conservative effects caused by damping and the applied dc magnetic field (rather than spin-transfer) may lead to the formation of magnetization self-oscillations. This suggests the intrinsic controllability of these oscillations by the applied dc magnetic field, a feature that may be potentially useful in the development of novel nano-magnetometers.

To discuss the essence of these phenomena, consider a single-domain nanomagnet with total (i.e., crystal + shape) ellipsoidal anisotropy and principal axes along $x,y,z$. The energy of the system can be written in dimensionless form as: $\gM( \mvec ) = \left ( D_x m_x^2 + D_y m_y^2 + D_z m_z^2 \right ) / 2$. In this expression, energy is measured in units of $\mu_0 \Msat^2 V$ ($V$ is the volume of the nanomagnet and $\Msat$ is the spontaneous magnetization), while $\mvec = (m_x,m_y,m_z)$ represents the normalized magnetization ( $|\mvec|^2 = 1$ ) of the nanomagnet. Assume that the $x$, $y$, and $z$ axes are the easy, intermediate, and hard anisotropy axes, respectively. The magnetic anisotropy coefficients are then ordered in the following manner: $D_x < D_y < D_z$. A typical case of interest is the disk-like free layer of a spin-transfer nanopillar device with in-plane anisotropy, for which $D_x < 0, D_y \simeq 0$, $D_z \simeq 1$. Under these conditions, it is convenient to shift the zero of energy by the amount $D_y/2$ and rewrite the energy as:

\beq
\gM( \mvec ) = \frac{1}{2} \left ( D_{zy} m_z^2 - D_{yx} m_x^2 \right ) \, \, \, ,
\label{EQ: gM}
\eeq

\noindent where $D_{yx} \equiv D_y - D_x > 0$, $D_{zy} \equiv D_z - D_y > 0$, and use has been made of the identity $m_x^2+m_y^2+m_z^2 = 1$.

Assume now that the nanomagnet is subjected to a spin-transfer torque of the form $\beta \, \mvec \times \left ( \mvec \times \evecy \right )$, due to a spin current with polarisation parallel to the intermediate axis $\evecy$. The dimensionless parameter $\beta$ measures the intensity of the spin current. The equation for the magnetization dynamics in the absence of intrinsic (thermal) damping is:

\beq
\frac{d \mvec}{d t} = - \mvec \times \hvecm + \beta \, \mvec \times \left ( \mvec \times \evecy \right ) \, \, \, ,
\label{EQ: LL}
\eeq

\noindent where $\hvecm \equiv - \partial \gM / \partial \mvec = D_{yx} m_x \evecx  - D_{zy} m_z \evecz$. Equation (\ref{EQ: LL}) is also dimensionless, with time measured in units of $\left ( \gamma \Msat \right )^{-1}$ ($\gamma$ is the absolute value of the gyromagnetic ratio). The dynamics preserves the magnitude of magnetization and thus takes place on the surface of the unit sphere $|\mvec|^2 = 1$. Of special importance are the two points: $m_y = \pm 1$, $m_x = m_z = 0$, which are critical points of the dynamics for any arbitrary value of the spin current.

The dynamics (\ref{EQ: LL}) is characterised by a conservation law. This follows from the fact that Eq. (\ref{EQ: LL}) is invariant under the transformation: $m_y \rightarrow - m_y$, $t \rightarrow -t$. To explain this, consider first the purely  precessional dynamics under zero current: $d \mvec / d t = - \mvec \times \hvecm$. The trajectories of this dynamics on the unit sphere $|\mvec|^2 = 1$ are constant-level lines of the anisotropy energy (\ref{EQ: gM}). When considered as a function of $\mvec$ on the unit sphere, this energy is an even function of $m_y$. Consequently, its maxima and minima lie on the plane $m_y = 0$ and all its constant-level curves intersect the plane $m_y = 0$ at least once.

As the spin current $\beta$ is gradually increased from $\beta = 0$, the property of these trajectories on the unit sphere of intersecting the plane $m_y = 0$ cannot be immediately destroyed by the current, because of continuity. Consider one of these trajectories, and choose the time origin at the moment when the plane $m_y = 0$ is crossed. Then, because of the mentioned $(m_y, t)$ reversal symmetry, the trajectory will consist of two parts, a forward-in-time and a backward-in-time parts, mirror-symmetric with respect to the plane $m_y = 0$. Consequently, if that trajectory crosses the plane $m_y = 0$ a second time, it is a closed trajectory. Trajectories with more than two intersections with $m_y = 0$ are not possible. If a trajectory is not closed, then, again because of the $(m_y,t)$ reversal symmetry, it necessarily connects two critical points characterized by opposite values of $m_y$. These critical points are the points $(m_x=0, m_y = \pm 1, m_z = 0)$, which are saddle points of the dynamics if the current is not too large.

Therefore, one arrives at the important conclusion that the phase portrait consists only of closed trajectories or open trajectories connecting saddle points (so-called separatrix trajectories) even under nonzero spin current. These closed trajectories and separatrices can be interpreted as constant-level curves of some conserved quantity \cite{Andronov1987}, say $\gtildeM( \mvec; \beta)$. Consequently, there must exist an integrating factor $f(\mvec; \beta)$ reducing Eq. (\ref{EQ: LL}) to the conservative form:

\beq
\frac{d \mvec}{d \, t} = \frac{1}{f} \, \mvec \times \frac{ \partial \gtildeM }{ \partial \mvec } \, \, \, ,
\label{EQ: LL gtildeM}
\eeq

\noindent where the conserved quantity is $\gtildeM(\mvec; \beta) = f(\mvec; \beta) \, \gM(\mvec)$ and $f(\mvec; 0) \equiv 1$, in order to guarantee that the dynamics is reduced to $d \mvec / d t = - \mvec \times \hvecm$ when no spin current is injected.

\begin{figure}[hbt]
\begin{tabular}{c}
\includegraphics[width=0.485\textwidth]{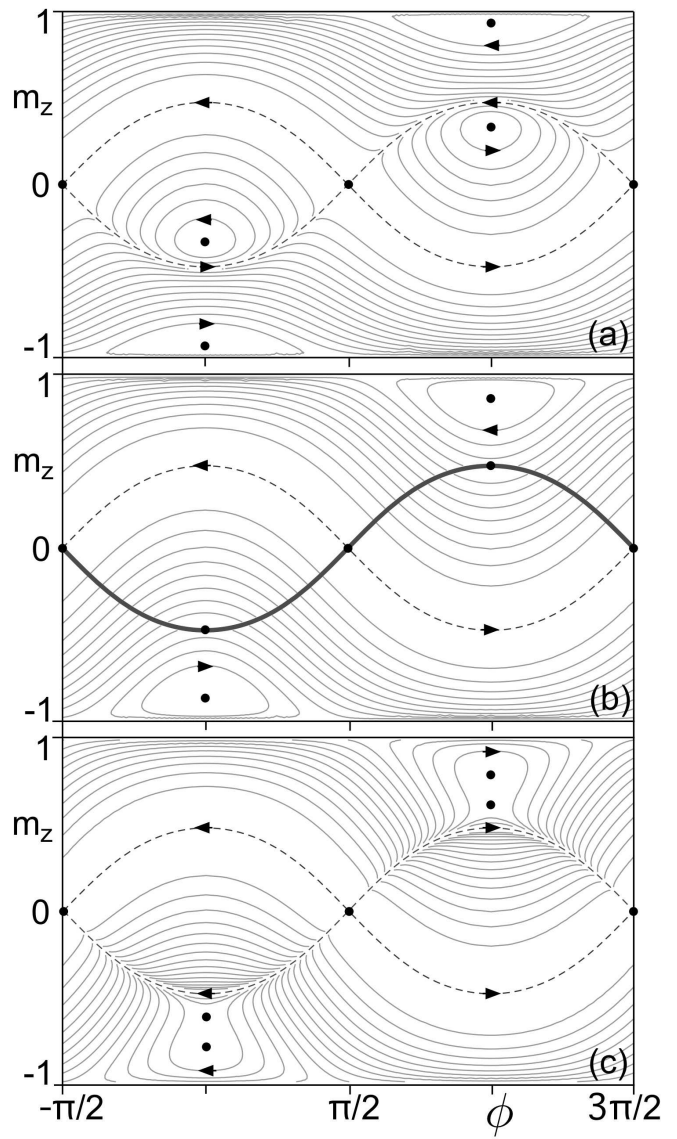}
\end{tabular}
\caption{\label{FIG: gtildeM} Phase portrait of undamped spin-transfer-driven dynamics under zero external field and increasing spin current. Continuous lines: constant-level curves of integral of motion $\gtildeM$. Arrows: direction of magnetization change. Dashed lines: invariant trajectories $w \pm w_Q = 0$. Black dots: critical points. (a) $\beta = 0.75 \, Q$. (b) $\beta = Q$ (bifurcation point), the bold continuous line representing the critical line on which $d \mvec / dt = 0$. (c) $\beta = 1.15 \, Q$. Parameters: $D_{yx} = 0.3$, $D_{zy} = 1$, $Q = \sqrt{D_{zy} D_{yx}}$. }
\end{figure}

To derive this integrating factor, consider that $\gtildeM$, as a conserved quantity, must satisfy the condition: $d \gtildeM / dt = (\partial \gtildeM / \partial \mvec) \cdot (d \mvec / dt) = 0$ along magnetization trajectories, that is (see Eq. (\ref{EQ: LL})):

\beq
\left ( \hvecm - \frac{\gM}{f} \, \frac{\partial f}{\partial \mvec} \right ) \cdot \Big [ \mvec \times \big ( \hvecm - \beta \, \mvec \times \evecy \big ) \Big ] = 0 \, \, \, .
\label{EQ: gtildeM cond}
\eeq

\noindent This condition is identically satisfied for any $\mvec$ if the vectors $\hvecm - \left ( \gM / f \right )\partial f / \partial \mvec$ and $\hvecm - \beta \, \mvec \times \evecy$ are parallel, that is, if:

\beq
\frac{\gM}{f} \, \frac{\partial f}{\partial \mvec} = \beta \, \mvec \times \evecy \, \, \, .
\label{EQ: dfdm}
\eeq

\noindent This equation yields the following differential equation:

\beq
\frac{1}{f} \, \frac{df}{dw} = \frac{\beta}{Q} \, \, \frac{2 w_Q}{w^2 - w_Q^2} \, \, \, , \, \, \, w = \frac{m_z}{m_x} \, \, \, ,
\label{EQ: df dw}
\eeq

\noindent where: $Q = \sqrt{D_{zy} D_{yx}}$ and $w_Q = \sqrt{D_{yx} / D_{zy}}$. Indeed, $\mvec \times \evecy = m_x^2 \, \partial w / \partial \mvec$ and $\gM (\mvec) = D_{zy} m_x^2 \left ( w^2 - w_Q^2 \right ) / 2$ (see Eq. (\ref{EQ: gM})). By integrating Eq. (\ref{EQ: df dw}) under the condition that $f(w; 0) \equiv 1$, one obtains:

\beq
f(w; \beta) = \left | \frac{w - w_Q}{w + w_Q} \right |^{\beta / Q} \, \, \, .
\label{EQ: f w}
\eeq

\noindent By using Eq. (\ref{EQ: dfdm}), Eq. (\ref{EQ: LL}) is transformed into Eq. (\ref{EQ: LL gtildeM}), as anticipated. The phase portrait of the dynamics is straightforwardly obtained by drawing the constant-level lines of $\gtildeM(\mvec;\beta) \equiv f(w;\beta) \, \gM(\mvec)$ (Fig. \ref{FIG: gtildeM}). A  convenient representation is obtained in terms of cylindrical coordinates $( m_z, \phi)$, whose relation to the cartesian magnetization components $(m_x, m_y, m_z)$ is: $m_x = \sqrt{1 - m_z^2} \, \cos \phi$, $m_y = \sqrt{1 - m_z^2} \, \sin \phi$.

Figure \ref{FIG: gtildeM} illustrates the progressive restructuring of the dynamics as the spin current is increased. A remarkable property of this restructuring is the invariance of zero-energy trajectories. According to Eq. (\ref{EQ: gM}), constant-level lines on which $\gM = 0$ are described by the equations: $m_z \sqrt{D_{zy}} \pm m_x \sqrt{D_{yx}} = 0$. It is easily verified that on these curves $\hvecm = \pm Q \, \mvec \times \evecy$. By substituting this expression into Eq. (\ref{EQ: LL}) and by taking into account that $\hvecm = - \partial \gM / \partial \mvec$, one obtains:

\beq
\frac{d \mvec}{d t} = \left ( 1 \mp \frac{\beta}{Q} \right ) \mvec \times \frac{\partial \gM}{\partial \mvec} \, \, \, .
\label{EQ: dmdt critical}
\eeq

\noindent This expression reveals that if $\gM = 0$, then $d \gM / dt = (\partial \gM / \partial \mvec) \cdot (d \mvec / dt) = 0$. In other words, constant-level curves on which $\gM = 0$ are always solutions of the dynamics, whatever the value of the spin current $\beta$. The current only affects the rate at which the constant-level curve is traversed. This rate goes to zero when $\beta = \pm Q$. When this condition is met, the entire constant-level curve becomes a critical line along which $d \mvec / dt = 0$. This occurs as a result of a complex bifurcation (see Fig. \ref{FIG: gtildeM}(b)), at which a global transition from closed to open magnetization trajectories occurs.

The dramatic restructuring of the phase portrait at $\beta = \pm Q$ affects the integral of motion $\gtildeM$, which is single-valued and continuous everywhere for $\beta^2 < Q^2$, while it diverges on the curve $w + w_Q = 0$ when $\beta > Q$ or on the curve $w - w_Q = 0$ when $\beta < -Q$. However, this divergence can be eliminated by taking advantage of the fact that any arbitrary monotone function $F(\gtildeM)$ can be taken as integral of motion instead of $\gtildeM$. If one chooses $F(\gtildeM) =\arctan \gtildeM$, then Eq. (\ref{EQ: LL gtildeM}) is transformed into:

\beq
\frac{d \mvec}{d \, t} = \frac{1 + \gtildeM^2}{f} \, \mvec \times \frac{ \partial}{ \partial \mvec } \arctan \gtildeM \, \, \, .
\label{EQ: LL gtildeM arctan}
\eeq

\noindent The function $\arctan \gtildeM$ appears similar to a stream function: it is conserved along magnetization trajectories and exhibits a discontinuous jump of amplitude equal to $\pi$ across the curve on which $\gtildeM$ diverges.

When a magnetic field $h_a \evecy$ is applied along the intermediate axis $\evecy$, the energy of the system becomes $g( \mvec; h_a ) = \gM(\mvec) - h_a \, \evecy \cdot \mvec$, and the undamped spin-transfer-driven dynamics is governed, instead of Eq. (\ref{EQ: LL gtildeM}), by the equation:

\beq
\frac{d \mvec}{d t} = \frac{1}{f} \, \mvec \times \frac{ \partial \gtildeM }{ \partial \mvec } - h_a \, \mvec \times \evecy \, \, \, .
\label{EQ: LL gtildeM h}
\eeq

\noindent The introduction of the field breaks the $(m_y,t)$ reversal symmetry and thus destroys the property of $\gtildeM$ of being an integral of motion. However, quite remarkably, it is possible to modify $\gtildeM$ in order to obtain a state function that acts as a global Lyapunov function \cite{Andronov1987,Perko1996}, that is, a function that monotonically increases or decreases under all circumstances during the magnetization process. We shall limit the discussion to the current interval $\beta^2 < Q^2$, in which $\gtildeM$ is a single-valued, well-behaved state function. A different approach, not discussed here, would be necessary to deal with the case when $\beta^2 > Q^2$.

\begin{figure}[hbt]
\begin{center}
\includegraphics[width=0.485\textwidth]{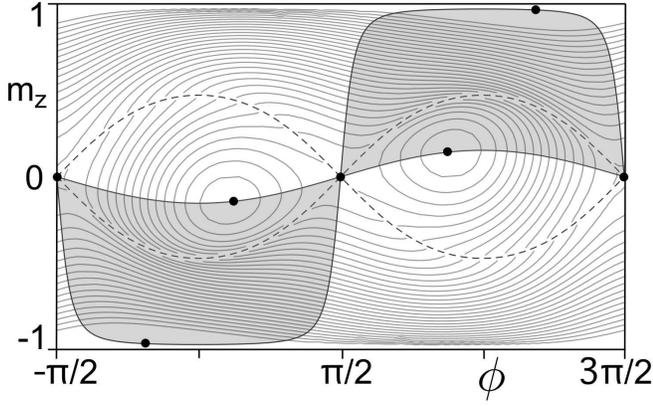}
\end{center}
\caption{\label{FIG: gtilde} Constant-level curves of function $\gtilde(\mvec;\beta, h_a)$ for $\beta^2 < Q^2$. Dashed lines: curves $w \pm w_Q = 0$ on which $\gM(\mvec) = 0$. Black dots: critical points. Shadowed region: region delimited by curves $w = w_1$ and  $w = w_2$, in which $( f - f_0 ) > 0$ and $(w - w_1)(w - w_2) < 0$ (the opposite occurs in remaining non shadowed region). Parameters: $D_{yx} = 0.3$, $D_{zy} = 1$, $\beta = 0.2$, $h_a = 0.1$, $w_Q = \sqrt{D_{zy} / D_{yx}}$. }
\end{figure}

The time derivative of $\gtildeM$ derived from Eq. (\ref{EQ: LL gtildeM h}) is: $d \gtildeM / dt = h_a \, f \, d m_y / d t$. The term $d m_y / d t$ can be computed from Eq. (\ref{EQ: LL gtildeM h}). One finds:

\beq
\frac{1}{1 - m_y^2} \, \frac{d m_y}{d t} = - \left ( \frac{2 R w}{1 + w^2} + \beta \right ) \, \, \, ,
\label{EQ: dmydt w}
\eeq

\noindent where $w = m_z / m_x$ and $R = \left ( D_{zy} + D_{yx} \right ) / 2$. Equation (\ref{EQ: dmydt w}) shows that the sign of $d m_y / d t$ is fully controlled by the roots of the equation $w^2 + 2 R w / \beta + 1 = 0$, namely, $w_{1,2} = - \left ( R \mp \sqrt{R^2 - \beta^2} \right ) / \beta$. When $\beta^2 < Q^2 < R^2$, $w_1^2 < w_Q^2$ and $w_2^2 > w_Q^2$. Therefore, the curve $w = w_1$ lies in the region $\gM < 0$ and the curve $w = w_2$ in the region $\gM > 0$, since $\gM (\mvec) = D_{zy} m_x^2 \left ( w^2 - w_Q^2 \right ) / 2$ (see Eq. (\ref{EQ: gM}) and Fig. \ref{FIG: gtilde}).

Consider now the function:

\bea
&& \gtilde \left ( \mvec ; \beta, h_a \right ) = \label{EQ: gtilde def} \\
&& \gtildeM(\mvec; \beta ) - f_0 h_a \evecy \cdot \mvec \, \, \, , \, \, \,
f_0 = \left \{ \begin{array}{c} f(w_1; \beta) \, \, \, \text{if} \, \, \, \gM \leq 0 \\ f(w_2; \beta) \, \, \, \text{if} \, \, \, \gM > 0 \end{array} \right . .
\nonumber
\eea

\noindent Its time derivative, computed from Eq. (\ref{EQ: LL gtildeM h}), is: $d \gtilde / dt = h_a \left ( f - f_0 \right ) d m_y / d t$, at every point in state space at which $\gM \neq 0$. By combining this result with Eq. (\ref{EQ: dmydt w}), one arrives at:

\beq
\frac{d \gtilde}{dt} = - \beta h_a \, \frac{1 - m_y^2}{1 + w^2} \, \big ( f - f_0 \big ) \big ( w - w_1 \big ) \big ( w - w_2 \big ) \, \, \, .
\label{EQ: dgtildedt w}
\eeq

\noindent By definition of $f(w; \beta)$ (Eq. (\ref{EQ: f w})) and $f_0$ (Eq. (\ref{EQ: gtilde def})), $( f - f_0 ) > 0$ when $(w - w_1)(w - w_2) < 0$ and vice versa (see Fig. \ref{FIG: gtilde}). Therefore, the function $\gtilde(\mvec; \beta, h_a)$ will be an increasing or decreasing function of time, depending on whether the product $\beta h_a$ is positive or negative, respectively. In particular, the maxima and minima of $\gtilde$ will represent critical points of the dynamics (Fig. \ref{FIG: gtilde}). When $\beta = 0$ or $h_a = 0$, $\gtilde$ is reduced to the corresponding conserved quantity, $g(\mvec; h_a)$ or $\gtildeM(\mvec; \beta)$, respectively.

Equation (\ref{EQ: dgtildedt w}) is not valid when $\gM = 0$, because $\gtilde$ is discontinuous there as a consequence of the jump in $f_0$. However, this discontinuity does not modify the conclusions of our analysis. It is sufficient to complement Eq. (\ref{EQ: dgtildedt w}) with the information about the direction of crossing of the boundary $\gM = 0$. This information is readily obtained from the dynamics of the ratio $w = m_z / m_x$. From Eq. (\ref{EQ: LL gtildeM h}) one finds that if $w \pm w_Q = 0$, then $d w / dt = - \left ( 1 + w_Q^2 \right ) h_a$. Thus, the boundary $\gM = 0$ is crossed in the sense of decreasing $w$ when $h_a > 0$ and increasing $w$ when $h_a < 0$.

\begin{figure}[hbt]
\begin{center}
\includegraphics[width=0.485\textwidth]{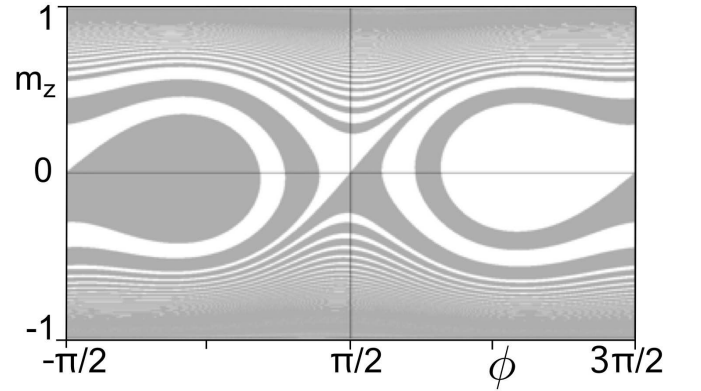}
\end{center}
\caption{\label{FIG: interlacing} Basins of attraction of undamped spin-transfer-driven dynamics under nonzero external magnetic field. Parameters: $D_{yx} = 0.3$, $D_{zy} = 1$, $\beta = 0.05$, $h_a = - 0.025$. }
\end{figure}

The existence of the function $\gtilde$ implies that the undamped spin-transfer-driven dynamics under nonzero field is nothing but a field-induced relaxation process toward $\gtilde$ minima or maxima, depending on the sign of the product $\beta h_a$. The function $\gtilde$ is characterised by a pair of minima in the region $\gM < 0$ and a pair of maxima in the region $\gM > 0$ \cite{Note01}. Therefore, there will exist two basins of attraction for the field-induced relaxation. These basins exhibit some degree of interlacing, similarly to what one observes in conventional magnetization relaxation due to intrinsic damping \cite{Bertotti2013}, the smaller the field $h_a$, the finer the interlacing. An example, obtained by numerical integration of Eq. (\ref{EQ: LL gtildeM h}), is shown in Fig. \ref{FIG: interlacing}. The fine basin interlacing makes the field-induced relaxation probabilistic in nature whenever control of initial conditions is imperfect \cite{Neishtadt1991,Arnold2006}. We stress that intrinsic damping has been neglected in the derivation of these results.

Intrinsic damping effects can be conveniently introduced in so-called Gilbert form \cite{Gilbert2004,Tserkovnyak2005,Hickey2009}, which amounts to changing $d \mvec / d t$ into $d \mvec / dt - \alpha \, \mvec \times d \mvec / dt$ in Eq. (\ref{EQ: LL gtildeM h}). As a consequence, Eq. (\ref{EQ: dgtildedt w}) is modified as:

\beq
\frac{d \gtilde}{dt} = - \alpha \, f \left | \frac{d \mvec}{d t} \right |^2 + h_a \left ( f - f_0 \right ) \frac{d m_y}{d t} \, \, \, ,
\label{EQ: dgtildedt alpha}
\eeq

\noindent where $\alpha$ represents the damping constant. The last term on the right-hand side of Eq. (\ref{EQ: dgtildedt alpha}) is not exactly coincident with the right-hand side of Eq. (\ref{EQ: dgtildedt w}), because $d m_y / dt$ is slightly modified by the introduction of damping. However, this modification plays a secondary role if $\alpha \ll 1$. In essence, Eq. (\ref{EQ: dgtildedt alpha}) is controlled by two terms, of which the one due to damping is always negative whereas the one due to current and field is approximately equal to the right-hand side of Eq. (\ref{EQ: dgtildedt w}) and has thus the sign of the product $\beta h_a$. Hence, when current and field have opposite sign, their action contributes jointly with intrinsic damping to the stabilization of $\gtilde$ minima, whereas when they have identical sign their action competes with that of intrinsic damping.

There exist conditions under which damping-induced and field-induced relaxations balance each other, leading to the appearance of limit cycles, that is, of magnetization self-oscillations. A typical scenario, confirmed by computer simulations, is the formation of a pair of attractive/repulsive limit cycles through a semi-stable limit-cycle bifurcation \cite{Kuznetsov1995,Perko1996}. Depending on the values of field and current, hysteretic transitions may occur between coexisting stationary and self-oscillation regimes, or conditions can be realised in which all critical points are unstable, which means that self-oscillations become the only possible steady-state regime available to the system.

This work was partially supported by Progetto Premiale MIUR-INRIM ÓNanotecnologie per la metrologia elettromagneticaÓ, by MIUR-PRIN Project 2010ECA8P3 ÓDyNanoMagÓ, and by NSF.
 

\begin{thebibliography}{24}
\expandafter\ifx\csname natexlab\endcsname\relax\def\natexlab#1{#1}\fi
\expandafter\ifx\csname bibnamefont\endcsname\relax
  \def\bibnamefont#1{#1}\fi
\expandafter\ifx\csname bibfnamefont\endcsname\relax
  \def\bibfnamefont#1{#1}\fi
\expandafter\ifx\csname citenamefont\endcsname\relax
  \def\citenamefont#1{#1}\fi
\expandafter\ifx\csname url\endcsname\relax
  \def\url#1{\texttt{#1}}\fi
\expandafter\ifx\csname urlprefix\endcsname\relax\def\urlprefix{URL }\fi
\providecommand{\bibinfo}[2]{#2}
\providecommand{\eprint}[2][]{\url{#2}}

\bibitem[{\citenamefont{Slonczewski}(1996)}]{Slonczewski1996}
\bibinfo{author}{\bibfnamefont{J.~C.} \bibnamefont{Slonczewski}},
  \bibinfo{journal}{J. Magn. Magn. Mater.} \textbf{\bibinfo{volume}{159}},
  \bibinfo{pages}{L1} (\bibinfo{year}{1996}).

\bibitem[{\citenamefont{Berger}(1996)}]{Berger1996}
\bibinfo{author}{\bibfnamefont{L.}~\bibnamefont{Berger}},
  \bibinfo{journal}{Phys. Rev. B} \textbf{\bibinfo{volume}{54}},
  \bibinfo{pages}{9353} (\bibinfo{year}{1996}).

\bibitem[{\citenamefont{Tserkovnyak et~al.}(2005)\citenamefont{Tserkovnyak,
  Brataas, Bauer, and Halperin}}]{Tserkovnyak2005}
\bibinfo{author}{\bibfnamefont{Y.}~\bibnamefont{Tserkovnyak}},
  \bibinfo{author}{\bibfnamefont{A.}~\bibnamefont{Brataas}},
  \bibinfo{author}{\bibfnamefont{G.~E.~W.} \bibnamefont{Bauer}},
  \bibnamefont{and} \bibinfo{author}{\bibfnamefont{B.~I.}
  \bibnamefont{Halperin}}, \bibinfo{journal}{Rev. Mod. Phys.}
  \textbf{\bibinfo{volume}{77}}, \bibinfo{pages}{1375} (\bibinfo{year}{2005}).

\bibitem[{\citenamefont{Haney et~al.}(2008)\citenamefont{Haney, Duine, Nunez,
  and MacDonald}}]{Haney2008}
\bibinfo{author}{\bibfnamefont{P.~M.} \bibnamefont{Haney}},
  \bibinfo{author}{\bibfnamefont{R.~A.} \bibnamefont{Duine}},
  \bibinfo{author}{\bibfnamefont{A.~S.} \bibnamefont{Nunez}}, \bibnamefont{and}
  \bibinfo{author}{\bibfnamefont{A.~H.} \bibnamefont{MacDonald}},
  \bibinfo{journal}{J. Magn. Magn. Mater.} \textbf{\bibinfo{volume}{320}},
  \bibinfo{pages}{1300} (\bibinfo{year}{2008}).

\bibitem[{\citenamefont{Brataas et~al.}(2012)\citenamefont{Brataas, Kent, and
  Ohno}}]{Brataas2012}
\bibinfo{author}{\bibfnamefont{A.}~\bibnamefont{Brataas}},
  \bibinfo{author}{\bibfnamefont{A.~D.} \bibnamefont{Kent}}, \bibnamefont{and}
  \bibinfo{author}{\bibfnamefont{H.}~\bibnamefont{Ohno}},
  \bibinfo{journal}{Nature Mater.} \textbf{\bibinfo{volume}{11}},
  \bibinfo{pages}{372} (\bibinfo{year}{2012}).

\bibitem[{\citenamefont{Katine et~al.}(2000)\citenamefont{Katine, Albert,
  Buhrman, Myers, and Ralph}}]{Katine2000}
\bibinfo{author}{\bibfnamefont{J.~A.} \bibnamefont{Katine}},
  \bibinfo{author}{\bibfnamefont{F.~J.} \bibnamefont{Albert}},
  \bibinfo{author}{\bibfnamefont{R.~A.} \bibnamefont{Buhrman}},
  \bibinfo{author}{\bibfnamefont{E.~B.} \bibnamefont{Myers}}, \bibnamefont{and}
  \bibinfo{author}{\bibfnamefont{D.~C.} \bibnamefont{Ralph}},
  \bibinfo{journal}{Phys. Rev. Lett.} \textbf{\bibinfo{volume}{84}},
  \bibinfo{pages}{3149} (\bibinfo{year}{2000}).

\bibitem[{\citenamefont{Bazaliy et~al.}(2004)\citenamefont{Bazaliy, Jones, and
  Zhang}}]{Bazaliy2004}
\bibinfo{author}{\bibfnamefont{Y.~B.} \bibnamefont{Bazaliy}},
  \bibinfo{author}{\bibfnamefont{B.~A.} \bibnamefont{Jones}}, \bibnamefont{and}
  \bibinfo{author}{\bibfnamefont{S.~C.} \bibnamefont{Zhang}},
  \bibinfo{journal}{Phys. Rev. B} \textbf{\bibinfo{volume}{69}},
  \bibinfo{pages}{094421} (\bibinfo{year}{2004}).

\bibitem[{\citenamefont{Kubota et~al.}(2008)\citenamefont{Kubota, Fukushima,
  Yakushiji, Nagahama, Yuasa, Ando, Maehara, Nagamine, Tsunekawa, Djayaprawira
  et~al.}}]{Kubota2008}
\bibinfo{author}{\bibfnamefont{H.}~\bibnamefont{Kubota}},
  \bibinfo{author}{\bibfnamefont{A.}~\bibnamefont{Fukushima}},
  \bibinfo{author}{\bibfnamefont{K.}~\bibnamefont{Yakushiji}},
  \bibinfo{author}{\bibfnamefont{T.}~\bibnamefont{Nagahama}},
  \bibinfo{author}{\bibfnamefont{S.}~\bibnamefont{Yuasa}},
  \bibinfo{author}{\bibfnamefont{K.}~\bibnamefont{Ando}},
  \bibinfo{author}{\bibfnamefont{H.}~\bibnamefont{Maehara}},
  \bibinfo{author}{\bibfnamefont{Y.}~\bibnamefont{Nagamine}},
  \bibinfo{author}{\bibfnamefont{K.}~\bibnamefont{Tsunekawa}},
  \bibinfo{author}{\bibfnamefont{D.~D.} \bibnamefont{Djayaprawira}},
  \bibnamefont{et~al.}, \bibinfo{journal}{Nature Physics}
  \textbf{\bibinfo{volume}{4}}, \bibinfo{pages}{37} (\bibinfo{year}{2008}).

\bibitem[{\citenamefont{Liu et~al.}(2012)\citenamefont{Liu, Pai, Li, Tseng,
  Ralph, and Buhrman}}]{Liu2012}
\bibinfo{author}{\bibfnamefont{L.}~\bibnamefont{Liu}},
  \bibinfo{author}{\bibfnamefont{C.~F.} \bibnamefont{Pai}},
  \bibinfo{author}{\bibfnamefont{Y.}~\bibnamefont{Li}},
  \bibinfo{author}{\bibfnamefont{H.~W.} \bibnamefont{Tseng}},
  \bibinfo{author}{\bibfnamefont{D.~C.} \bibnamefont{Ralph}}, \bibnamefont{and}
  \bibinfo{author}{\bibfnamefont{R.~A.} \bibnamefont{Buhrman}},
  \bibinfo{journal}{Science} \textbf{\bibinfo{volume}{336}},
  \bibinfo{pages}{555} (\bibinfo{year}{2012}).

\bibitem[{\citenamefont{Kiselev et~al.}(2003)\citenamefont{Kiselev, Sankey,
  Krivorotov, Emley, Schoelkopf, Buhrman, and Ralph}}]{Kiselev2003}
\bibinfo{author}{\bibfnamefont{S.~I.} \bibnamefont{Kiselev}},
  \bibinfo{author}{\bibfnamefont{J.~C.} \bibnamefont{Sankey}},
  \bibinfo{author}{\bibfnamefont{I.~N.} \bibnamefont{Krivorotov}},
  \bibinfo{author}{\bibfnamefont{N.~C.} \bibnamefont{Emley}},
  \bibinfo{author}{\bibfnamefont{R.~J.} \bibnamefont{Schoelkopf}},
  \bibinfo{author}{\bibfnamefont{R.~A.} \bibnamefont{Buhrman}},
  \bibnamefont{and} \bibinfo{author}{\bibfnamefont{D.~C.} \bibnamefont{Ralph}},
  \bibinfo{journal}{Nature} \textbf{\bibinfo{volume}{425}},
  \bibinfo{pages}{380} (\bibinfo{year}{2003}).

\bibitem[{\citenamefont{Rippard et~al.}(2004)\citenamefont{Rippard, Pufall,
  Kaka, Russek, and Silva}}]{Rippard2004}
\bibinfo{author}{\bibfnamefont{W.~H.} \bibnamefont{Rippard}},
  \bibinfo{author}{\bibfnamefont{M.~R.} \bibnamefont{Pufall}},
  \bibinfo{author}{\bibfnamefont{S.}~\bibnamefont{Kaka}},
  \bibinfo{author}{\bibfnamefont{S.~E.} \bibnamefont{Russek}},
  \bibnamefont{and} \bibinfo{author}{\bibfnamefont{T.~J.} \bibnamefont{Silva}},
  \bibinfo{journal}{Phys. Rev. Lett.} \textbf{\bibinfo{volume}{92}},
  \bibinfo{pages}{027201} (\bibinfo{year}{2004}).

\bibitem[{\citenamefont{Krivorotov et~al.}(2005)\citenamefont{Krivorotov,
  Emley, Sankey, Kiselev, Ralph, and Buhrman}}]{Krivorotov2005}
\bibinfo{author}{\bibfnamefont{I.~N.} \bibnamefont{Krivorotov}},
  \bibinfo{author}{\bibfnamefont{N.~C.} \bibnamefont{Emley}},
  \bibinfo{author}{\bibfnamefont{J.~C.} \bibnamefont{Sankey}},
  \bibinfo{author}{\bibfnamefont{S.~I.} \bibnamefont{Kiselev}},
  \bibinfo{author}{\bibfnamefont{D.~C.} \bibnamefont{Ralph}}, \bibnamefont{and}
  \bibinfo{author}{\bibfnamefont{R.~A.} \bibnamefont{Buhrman}},
  \bibinfo{journal}{Science} \textbf{\bibinfo{volume}{307}},
  \bibinfo{pages}{228} (\bibinfo{year}{2005}).

\bibitem[{\citenamefont{Bertotti et~al.}(2005)\citenamefont{Bertotti, Serpico,
  Mayergoyz, Magni, d'Aquino, and Bonin}}]{Bertotti2005}
\bibinfo{author}{\bibfnamefont{G.}~\bibnamefont{Bertotti}},
  \bibinfo{author}{\bibfnamefont{C.}~\bibnamefont{Serpico}},
  \bibinfo{author}{\bibfnamefont{I.~D.} \bibnamefont{Mayergoyz}},
  \bibinfo{author}{\bibfnamefont{A.}~\bibnamefont{Magni}},
  \bibinfo{author}{\bibfnamefont{M.}~\bibnamefont{d'Aquino}}, \bibnamefont{and}
  \bibinfo{author}{\bibfnamefont{R.}~\bibnamefont{Bonin}},
  \bibinfo{journal}{Phys. Rev. Lett.} \textbf{\bibinfo{volume}{94}},
  \bibinfo{pages}{127206} (\bibinfo{year}{2005}).

\bibitem[{\citenamefont{Bertotti et~al.}(2009)\citenamefont{Bertotti,
  Mayergoyz, and Serpico}}]{Bertotti2009}
\bibinfo{author}{\bibfnamefont{G.}~\bibnamefont{Bertotti}},
  \bibinfo{author}{\bibfnamefont{I.~D.} \bibnamefont{Mayergoyz}},
  \bibnamefont{and} \bibinfo{author}{\bibfnamefont{C.}~\bibnamefont{Serpico}},
  \emph{\bibinfo{title}{Nonlinear Magnetization Dynamics in Nanosystems}}
  (\bibinfo{publisher}{Elsevier}, \bibinfo{address}{Oxford},
  \bibinfo{year}{2009}).

\bibitem[{\citenamefont{Li et~al.}(2005)\citenamefont{Li, He, and
  Zhang}}]{Li2005}
\bibinfo{author}{\bibfnamefont{Z.}~\bibnamefont{Li}},
  \bibinfo{author}{\bibfnamefont{J.}~\bibnamefont{He}}, \bibnamefont{and}
  \bibinfo{author}{\bibfnamefont{S.}~\bibnamefont{Zhang}},
  \bibinfo{journal}{Phys. Rev. B} \textbf{\bibinfo{volume}{72}},
  \bibinfo{pages}{212411} (\bibinfo{year}{2005}).

\bibitem[{\citenamefont{Andronov et~al.}(1987)\citenamefont{Andronov, Vitt, and
  Khaikin}}]{Andronov1987}
\bibinfo{author}{\bibfnamefont{A.~A.} \bibnamefont{Andronov}},
  \bibinfo{author}{\bibfnamefont{A.~A.} \bibnamefont{Vitt}}, \bibnamefont{and}
  \bibinfo{author}{\bibfnamefont{S.~E.} \bibnamefont{Khaikin}},
  \emph{\bibinfo{title}{Theory of Oscillators}} (\bibinfo{publisher}{Dover},
  \bibinfo{address}{New York}, \bibinfo{year}{1987}).

\bibitem[{\citenamefont{Perko}(1996)}]{Perko1996}
\bibinfo{author}{\bibfnamefont{L.}~\bibnamefont{Perko}},
  \emph{\bibinfo{title}{Differential Equations and Dynamical Systems}}
  (\bibinfo{publisher}{Springer}, \bibinfo{address}{New York},
  \bibinfo{year}{1996}).

\bibitem[{Not()}]{Note01}
\bibinfo{note}{A detailed analysis, not reported here, shows that this occurs
  in the current-field region: $(h_a \pm h_0)^2 + \beta^2 < R^2$, where $h_0 =
  (D_{zy} - D_{yx})/2$ and $R = (D_{zy} + D_{yx})/2$.}

\bibitem[{\citenamefont{Bertotti et~al.}(2013)\citenamefont{Bertotti, Serpico,
  and Mayergoyz}}]{Bertotti2013}
\bibinfo{author}{\bibfnamefont{G.}~\bibnamefont{Bertotti}},
  \bibinfo{author}{\bibfnamefont{C.}~\bibnamefont{Serpico}}, \bibnamefont{and}
  \bibinfo{author}{\bibfnamefont{I.~D.} \bibnamefont{Mayergoyz}},
  \bibinfo{journal}{Phys. Rev. Lett.} \textbf{\bibinfo{volume}{110}},
  \bibinfo{pages}{147205} (\bibinfo{year}{2013}).

\bibitem[{\citenamefont{Neishtadt}(1991)}]{Neishtadt1991}
\bibinfo{author}{\bibfnamefont{A.~I.} \bibnamefont{Neishtadt}},
  \bibinfo{journal}{Chaos} \textbf{\bibinfo{volume}{1}}, \bibinfo{pages}{42}
  (\bibinfo{year}{1991}).

\bibitem[{\citenamefont{Arnold et~al.}(2006)\citenamefont{Arnold, Kozlov, and
  Neishtadt}}]{Arnold2006}
\bibinfo{author}{\bibfnamefont{V.~I.} \bibnamefont{Arnold}},
  \bibinfo{author}{\bibfnamefont{V.~V.} \bibnamefont{Kozlov}},
  \bibnamefont{and} \bibinfo{author}{\bibfnamefont{A.~I.}
  \bibnamefont{Neishtadt}}, \emph{\bibinfo{title}{Mathematical Aspects of
  Classical and Celestial Mechanics}} (\bibinfo{publisher}{Springer},
  \bibinfo{address}{Berlin}, \bibinfo{year}{2006}).

\bibitem[{\citenamefont{Gilbert}(2004)}]{Gilbert2004}
\bibinfo{author}{\bibfnamefont{T.~L.} \bibnamefont{Gilbert}},
  \bibinfo{journal}{IEEE Trans. Magn.} \textbf{\bibinfo{volume}{40}},
  \bibinfo{pages}{3443} (\bibinfo{year}{2004}).

\bibitem[{\citenamefont{Hickey and Moodera}(2009)}]{Hickey2009}
\bibinfo{author}{\bibfnamefont{M.~C.} \bibnamefont{Hickey}} \bibnamefont{and}
  \bibinfo{author}{\bibfnamefont{J.~S.} \bibnamefont{Moodera}},
  \bibinfo{journal}{Phys. Rev. Lett.} \textbf{\bibinfo{volume}{102}},
  \bibinfo{pages}{137601} (\bibinfo{year}{2009}).

\bibitem[{\citenamefont{Kuznetsov}(1995)}]{Kuznetsov1995}
\bibinfo{author}{\bibfnamefont{Y.~A.} \bibnamefont{Kuznetsov}},
  \emph{\bibinfo{title}{Elements of Applied Bifurcation Theory}}
  (\bibinfo{publisher}{Springer}, \bibinfo{address}{New York},
  \bibinfo{year}{1995}).

\end{thebibliography}


\end{document}